\documentclass{ws-procs9x6}
\usepackage{epsfig}
\newcommand{\vsig}{\mbox{\boldmath$\sigma$\unboldmath}}
\newcommand{\veps}{\mbox{\boldmath$\epsilon$\unboldmath}}

\newcommand{\be}{\begin{equation}}
\newcommand{\ee}{\end{equation}}
\newcommand{\bea}{\begin{eqnarray}}
\newcommand{\eea}{\end{eqnarray}}
\newcommand{\bean}{\begin{eqnarray*}}
\newcommand{\eean}{\end{eqnarray*}}

\newcommand{\gapproxeq}{\lower
.7ex\hbox{$\;\stackrel{\textstyle >}{\sim}\;$}}
\newcommand{\lapproxeq}{\lower
.7ex\hbox{$\;\stackrel{\textstyle <}{\sim}\;$}}

\begin{document}

\title{Study the exotic $\Theta^+$ in polarized photoproduction 
reactions
}

\author{Qiang Zhao
}

\address{Department of Physics\\
University of Surrey\\ 
Guildford, Surrey GU2 7XH, United Kingdom\\ 
E-mail: qiang.zhao@surrey.ac.uk}

\maketitle

\abstracts{
We present an analysis of a Beam-Target double polarization
asymmetry in $\gamma n\to \Theta^+ K^-$.
We show that this quantity can serve as a filter for the determination
of the $\Theta^+$'s spin-parity assignment near threshold.  It is
highly selective between $1/2^+$ and $1/2^-$ configurations due to
dynamical reasons. 
}

\section{Introduction}

The report of signals of a strangeness $S=+1$ baryon 
$\Theta^+$~\cite{theta-exp} 
has stimulated a tremendous number of activities 
in both experiment and theory in the past 12 months. 
Baryon with strangeness $S=+1$ and strongly coupled into $\bar{K} N$
implies that its minimum number of constituents
should be at least five, i.e. a pentaquark state $(uudd\bar{s})$. 
With a rather-low mass of 1.54 GeV and a narrow width of less than 25 MeV, 
the existence of such a state 
(later another state $\Xi^{--}$ was also reported~\cite{na49}) also initiates 
a tremedous number of 
questions about strong QCD dynamics in the low energy regime.  
Whether it is the chiral-soliton-model-predicted $\bar{\bf 10}$
multiplets~\cite{dpp}, 
or pentaquark $\bar{\bf 10}$ in the quark model~\cite{antidecuplet,close1}, 
such a state reveals some important new aspects of QCD, which have not been 
widely concerned before.

The quantum numbers of the $\Theta^+$ are undoubtedly one of the 
key issues for answering some of those critical questions. 
So far, the spin and parity of the $\Theta^+$ have not been well-determined
in experiment~\cite{nakano-nstar,burkert-nstar}, 
in spite of that the absence of signals in 
$\gamma p\to \Theta^{++}K^-$ favors $\Theta^+$ 
to be an isopsin singlet (see e.g. SAPHIR results).
Meanwhile, a series of theoretical approaches have been 
proposed to understand the nature of the $\Theta^+$.
Due to the limited space here, we will concentrate on the theoretical 
study of the $\Theta^+$ photoproduction on the nucleon.
A number of phenomenological approaches  have been proposed 
in the literature focussing on 
cross section predictions~\cite{reaction,nam}.  
However, due to the lack
of knowledge about the underlying reaction mechanism, such
studies of the reaction cross sections are strongly
model-dependent.  For instance, the total width of $\Theta^+$ still has
large uncertainties and could be much
narrower~\cite{nk-scatter,narrow-width,cckn}, 
and the role played by
$K^*$ exchange, as well as other {\it s}- and {\it u}-channel processes
are unknown. Also, in a phenomenological approach the energy dependence of the
couplings is generally introduced into the model via empirical form
factors, which will bring further uncertainties. Taking this into account, 
there are advantages with
polarization observables (e.g.  ambiguities arising from the unknown
form factors can be partially avoided). In association with the cross
section studies, supplementary information about the $\Theta^+$ can be
obtained~\cite{zhao-theta,za,nakayama,okl-2,rekalo}.

\section{A minimum phenomenology}

The following effective Lagrangians are assumed for the $\Theta N K$ couplings
with the spin-parity $1/2^+$ and $1/2^-$ for the $\Theta^+$, respectively:
\bea
\label{lagrangian}
L_{eff}(1/2^+) &=& g_{\Theta NK} 
\bar{\Theta} \gamma_\mu\gamma_5 \partial^\mu K N  + \mbox{h.c.},
\nonumber\\
L_{eff}(1/2^-) &=& g_{\Theta NK} \bar{\Theta} N K + \mbox{h.c.},
\eea
where $\bar{\Theta}$, $N$ and $K$ denote the field of $\Theta^+$, neutron
and $K^-$. 
The coupling constant $g_{\Theta NK}$ is determined 
by the experimental data for the decay width of $\Theta^+\to NK$, 
which however is still very imprecise. 
In this work, the same set of parameters as Refs.~\cite{zhao-theta,za}
is adopted by assuming $\Gamma_{\Theta^+\to NK}\simeq 10$ MeV. 
In Ref.~\cite{cz}, a narrower width of 1 MeV is adopted, based on
analyses of $N$-$K$ scattering data~\cite{nk-scatter}.
The change of such an overall factor will not change the 
behavior of polarization asymmetries for exclusive Born terms.
The magnetic moment of $\Theta^+$ is estimated 
based on the phenomenology of diquark model of Ref.~\cite{antidecuplet}, 
which is consistent with other models 
in the literature~\cite{magnet,nam}.

We also include the $K^*$ exchange in this model
as the leading contribution in association with the Born terms.
The $K^*N\Theta$ interaction is given by
\be
L_{\Theta N K^*}(1/2^+)=g_{\Theta N K^*}\bar{\Theta}
(\gamma_\mu +\frac{\kappa_\theta^*}{2M_\Theta}
\sigma_{\mu\nu}\partial^\nu)V^\mu N   + \mbox{h.c.} \ ,
\ee
and 
\be 
L_{\Theta N K^*}(1/2^-)=g_{\Theta N K^*}\bar{\Theta}
\gamma_5(\gamma_\mu +\frac{\kappa_\theta^*}{2M_\Theta}
\sigma_{\mu\nu}\partial^\nu)V^\mu N   + \mbox{h.c.} \ ,
\ee
where $g_{\Theta N K^*}$ and $\kappa_\theta^*$ denote
the vector and ``anomalous moment" couplings, respectively.
We follow Refs.~\cite{zhao-theta,za} to adopt the values for 
$g_{\Theta N K^*}$ and $\kappa_\theta^*$. 
It should be note that $|g_{\Theta N K^*}|=|g_{\Theta NK}|$ 
and $\kappa_\theta^*=0$
are widely adopted in the literature. 
In Ref.~\cite{cz}, $g^2_{\Theta N K^*}(1/2^+)=3g^2_{\Theta NK}$
and $\kappa_\theta^*=0$ are adopted in the calculation of 
cross sections for $1/2^+$ pentaquarks with the consideration of
the ``fall-apart" mechanism~\cite{JM,cckn,bs,cd}. The change of parameters
for the $K^*$ exchange does not change the behavior of the 
BT asymmetries for $\gamma n\to \Theta^+ K^-$ near threshold 
due to the importance of the Born terms.

The effective Lagrangian for the $K^*K\gamma$ vertex is given by
\be
L_{K^*K\gamma}=\frac{ie_0g_{K^*K\gamma}}{M_K}
\epsilon_{\alpha\beta\gamma\delta}
\partial^\alpha A^\beta\partial^\gamma V^\delta K  + \mbox{h.c.} \ ,
\ee
where $V^\delta$ denotes the $K^*$ field.
The coupling $g_{K^{*0}K^0\gamma}=1.13$ is determined
by $\Gamma_{K^{*0}\to K^0\gamma}=117$ keV~\cite{pdg2000}.

In the photoproduction reaction, defining 
the $z$-axis as the photon momentum direction, 
and the reaction plane in $x$-$z$ in the c.m. system of $\gamma$-$n$, 
the transition amplitude for $\gamma n\to K^-\Theta^+$
can be expressed as 
\be
\label{heli-amp}
T_{\lambda_\theta,\lambda_\gamma \lambda_N}\equiv
\langle \Theta^+, \lambda_\theta, {\bf P}_\theta; K^-, \lambda_0, {\bf q} 
| \hat{T} | n, \lambda_N, {\bf P}_i; \gamma, \lambda_\gamma, {\bf k}\rangle ,
\ee
where $\lambda_\gamma=\pm 1$, $\lambda_N\pm 1/2$, 
$\lambda_0=0$, and $\lambda_\theta$ 
are helicities 
of photon, neutron, $K^-$, and $\Theta^+$, respectively.

\section{Double polarization asymmetries}

We are interested in a Beam-Target (BT) double polarization
asymmetry measured in $\gamma n\to \Theta^+ K^-$. 
In this reaction
the photons are circularly polarized along 
the photon moment direction $\hat{z}$ and the neutrons transversely 
polarized along the $\hat{x}$-axis within the reaction plane.
In terms of the density matrix elements for the $\Theta^+$ decays, 
the BT asymmetry can be expressed as,
\be
\label{bt-heli}
D_{xz}=\frac{\rho^{BT}_{\frac 12,\frac 12}}{\rho^0_{\frac 12,\frac 12}} ,
\ee
where 
the subscript $xz$ denotes the polarization direction of
the the initial neutron target along $x$-axis in the production plane
and the incident photon along the $z$-axis, and 
the definition of the density matrix element is 
\be
\rho^{BT}_{\lambda_\theta\lambda_\theta^\prime}
=\frac{1}{2N}\sum_{\lambda_\gamma, \lambda_N}
\lambda_\gamma 
T_{\lambda_\theta,\lambda_\gamma -\lambda_N} 
T^*_{\lambda_\theta^\prime,\lambda_\gamma \lambda_N}  ,
\ee
where, $N=\frac 12 \sum_{\lambda_\theta,\lambda_\gamma,\lambda_N}
|T_{\lambda_\theta,\lambda_\gamma \lambda_N}|^2$ is the normalization
factor. 

As follows, we express the transition amplitudes 
in terms of the CGLN amplitudes~\cite{CGLN}. This is useful for 
our understanding the behavior of the BT asymmetry near threshold.

For $1/2^+$, we have
\be
\langle \Theta^+, \lambda_\theta, {\bf P}_\theta; K^-, \lambda_0, {\bf q} 
| \hat{T} | n, \lambda_N, {\bf P}_i; \gamma, \lambda_\gamma, {\bf k}\rangle 
=\langle \lambda_\theta | {\bf J}\cdot \veps_\gamma |\lambda_N \rangle  \ ,
\ee
with 
\be
\label{heli-posi}
{\bf J}\cdot \veps_\gamma= if_1\vsig \cdot \veps_\gamma
+ f_2\frac{1}{|{\bf q}||{\bf k}|}\vsig\cdot{\bf q}
\vsig\cdot({\bf k}\times\veps_\gamma)
+if_3\frac{1}{|{\bf q}||{\bf k}|}\vsig\cdot{\bf k}{\bf q}\cdot \veps_\gamma
+if_4\frac{1}{|{\bf q}|^2}\vsig\cdot{\bf q}{\bf q}\cdot \veps_\gamma .
\ee
The coefficients $f_{1,2,3,4}$ are functions of energies, momenta, 
and scattering angle $\theta_{c.m.}$, and contain information on dynamics.
They provide an alternative expression for the BT asymmetry:
\be
\label{bt-posi}
D_{xz}=\sin\theta_{c.m.}\mbox{Re}\{
f_1 f_3^* - f_2 f_4^* 
+ \cos\theta_{c.m.} (f_1 f_4^* -f_2 f_3^*)\} .
\ee

We are interested in the energy region near threshold,
namely, with the overall c.m. energy $W \sim 2.1$ GeV.
This is the region that 
the well-established 
contact term (Kroll-Ruderman term) dominates 
over all the other processes, and is the main component 
of $f_1$. 
The term of $f_2$ will have contributions 
from the {\it s}- and {\it u}-channel, while
the terms of $f_3$ and $f_4$ from the {\it t}- and {\it u}-channel. 
Near threshold, it shows that the {\it t}-channel amplitudes
have the least suppressions from the propagator $1/(t-M_K^2)$
in comparison with the {\it s}- and {\it u}-channel.
Also, the terms contributing to $f_4$ will be further 
suppressed near threshold due to the small momentum $|{\bf q}|$
in the final state. Considering the products of $f$ coefficients
in Eq.~(\ref{bt-posi}), we find the dominant contributions are from:
\be
\label{approx-1}
f_1 f_3^* \simeq 
-e_0^2g_{\Theta NK}^2
\frac{2}{t-M_K^2}
F_c(k, q)F_t(k, q) \ ,
\ee
where $F_c(k, q)$ and $F_t(k, q)$ are form factors
for the contact and {\it t}-channel, and are treated the same 
in this approach.

This kinematic analysis leads to the robust prediction of 
the behaviour of $D_{xz}$ near threshold to be 
dominated by 
\be 
D_{xz}\simeq\sin\theta_{c.m.}
\mbox{Re}\{ f_1 f_3^*\} .
\ee
Since the CGLN coefficients only depend weakly on 
$\theta_{c.m.}$ (via the Mandelstam variables), 
the dominance of the above term   
implies a $\sin\theta_{c.m.}$ behavior 
of $D_{xz}$, and the sign of $D_{xz}$ is determined by 
the product. 

\begin{figure}[htbp]
\begin{minipage}[tl]{54mm}
\epsfig{file=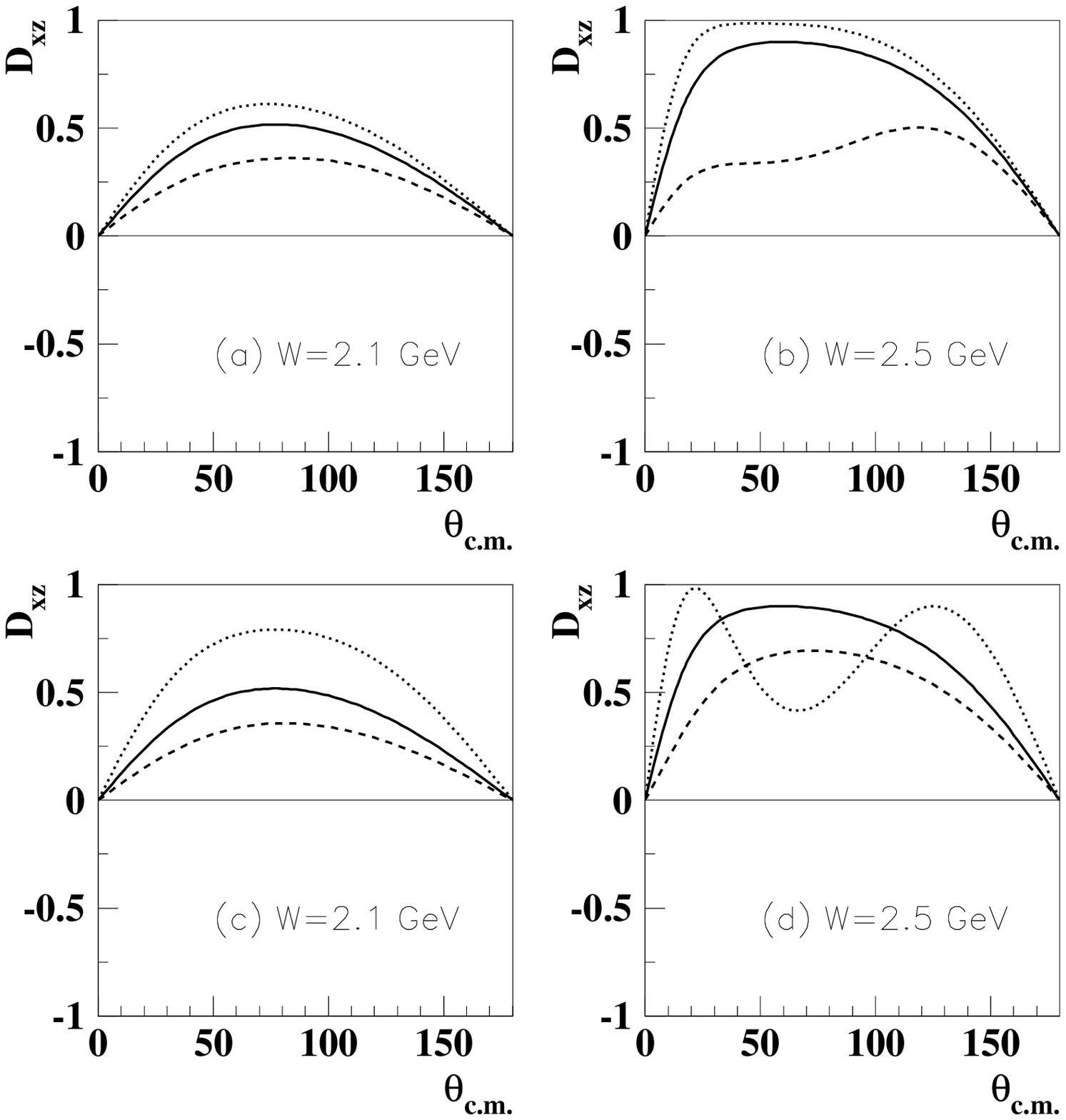, width=6cm,height=6.cm}
\caption{BT asymmetry for 
$\Theta^+$ of $1/2^+$ at $W=2.1$ and 2.5 GeV. 
The solid curves are results in the Born limit, while the dashed and
dotted curves denote results with the $K^*$ exchange included 
with different phases: $(g_{\Theta N K^*}, \kappa_\theta^*)= (-2.8,-3.71)$ 
(dashed curves in (a) and (b)),
$(+2.8,+3.71)$ (dotted curves in (a) and (b)), 
$(-2.8, +3.71)$ (dashed curves in (c) and (d)), and 
$(+2.8, -3.71)$ (dotted curves in (c) and (d)).}
\protect\label{fig:(2)}
\end{minipage}
%
%
\hspace{\fill}
%
\begin{minipage}[tl]{54mm}
\epsfig{file=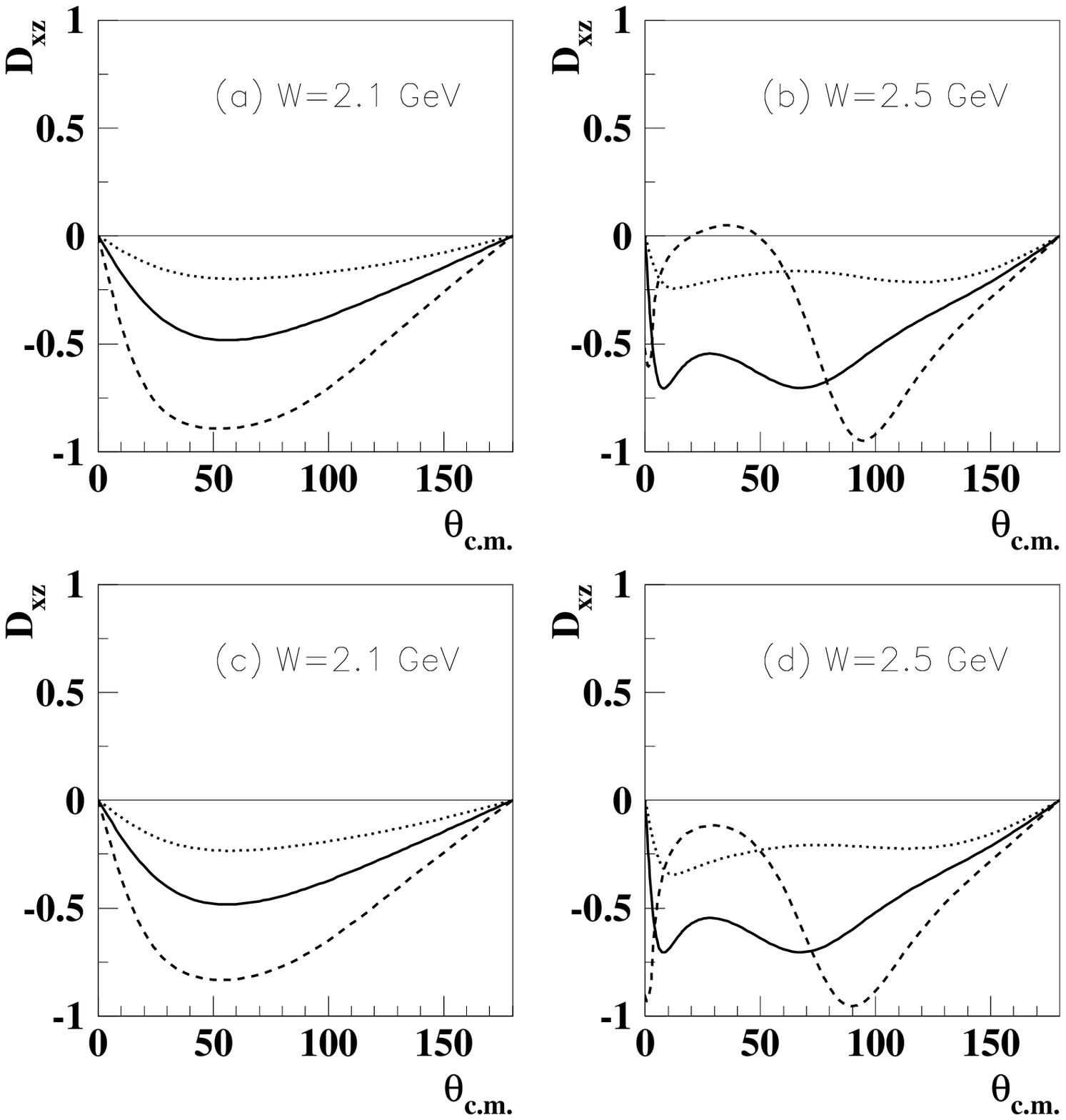, width=6cm,height=6.cm}
\caption{BT asymmetry for 
$\Theta^+$ of $1/2^-$ at $W=2.1$ and 2.5 GeV. 
The solid curves are results in the Born limit, while the dashed and
dotted curves denote results with the $K^*$ exchange included 
with different phases: $(g_{\Theta N K^*}, \kappa_\theta^*)= (-0.61,-0.371)$ 
(dashed curves in (a) and (b)),
$(+0.61,+0.371)$ (dotted curves in (a) and (b)), 
$(-0.61, +0.371)$ (dashed curves in (c) and (d)), and 
$(+0.61, -0.371)$ (dotted curves in (c) and (d)).}
\protect\label{fig:(3)}
\end{minipage}
%
\end{figure}

In Fig.~\ref{fig:(2)}, the numerical results 
for the BT asymmetry at $W=2.1$ GeV are presented.
The solid curve denote results in the Born limit, 
while the dashed and dotted curves represent results 
for including the $K^*$ exchange.
Although the sign change of the  $K^*$ exchange results in 
a quite significant change to the asymmetry values,
in all the cases, 
a clear $\sin\theta_{c.m.}$ behaviour appears in the BT asymmetry.
In particular, it is the contact and {\it t}-channel kaon exchange
in $f_1$ and $f_3$ that control the sign of $f_1 f_3^*$, and produce 
the positive BT asymmetry near threshold~\cite{za}. 
Also, note that it is natural that 
structures deviating from 
$\sin\theta_{c.m.}$ may arise at higher energies,
since other mechanisms could become important, and $f_1$ and $f_3$ will no longer 
be the leading terms. Such a feature can be seen 
through the asymmetries for different 
$K^*$ exchange phases at $W=2.5$ GeV, 
where the interference of 
the $\cos\theta_{c.m.}(f_1 f_4^*-f_2 f_3^*)$ term in Eq.~(\ref{bt-posi})
shows up indeed.

Similar analysis can be applied to the production of $1/2^-$. 
In general, the transition amplitudes can be arranged
in a way similar to the CGLN amplitudes for $1/2^+$:
\bea
{\bf J}\cdot\veps_\gamma &=&
iC_1\frac{1}{|{\bf k}|}\vsig\cdot(\veps_\gamma\times{\bf k})
+C_2\frac{1}{|{\bf q}|}\vsig\cdot{\bf q}\vsig\cdot\veps_\gamma\nonumber\\
&+ & iC_3\frac{1}{|{\bf q}||{\bf k}|^2}\vsig\cdot{\bf k}
{\bf q}\cdot (\veps_\gamma\times{\bf k})
+iC_4\frac{1}{|{\bf q}|^2|{\bf k}|}\vsig\cdot{\bf q}
{\bf q}\cdot (\veps_\gamma\times{\bf k}) ,
\eea
where coefficients $C_{1,2,3,4}$ are functions of energies, momenta and 
scattering angle, and contain dynamical information on the transitions. 
Restricted to the kinematics near threshold, 
a term proportional to 
${\bf q}\cdot\veps_\gamma\vsig\cdot({\bf k}\times{\bf q})$ 
in the {\it u}-channel is neglected in the above 
expression, but included in the calculation. 
Such a term is the same order of $C_4$. However, they
are both relatively suppressed in comparison with other terms 
due to the small $|{\bf q}|$ near threshold.
As follows, we neglect the term of 
${\bf q}\cdot\veps_\gamma\vsig\cdot({\bf k}\times{\bf q})$ 
and express the above in parallel to the CGLN amplitudes.

Since  $(\veps_\gamma\times\hat{\bf k})= i\lambda_\gamma\veps_\gamma$, 
one can replace vector $(\veps_\gamma\times\hat{\bf k})$
with $ i\lambda_\gamma\veps_\gamma$, and rewite the operator as
\bea
{\bf J}\cdot\veps_\gamma &=& 
i\lambda_\gamma\left[iC_1\vsig \cdot \veps_\gamma
+ C_2\frac{1}{|{\bf q}||{\bf k}|}\vsig\cdot{\bf q}
\vsig\cdot({\bf k}\times\veps_\gamma) \right.\nonumber\\
& & \left. +iC_3\frac{1}{|{\bf q}||{\bf k}|}
\vsig\cdot{\bf k}{\bf q}\cdot \veps_\gamma
+iC_4\frac{1}{|{\bf q}|^2}\vsig\cdot{\bf q}{\bf q}\cdot \veps_\gamma \right],
\eea
which has exactly the same form as Eq.~(\ref{heli-posi}) apart from an overall
phase factor from the photon polarization $i\lambda_\gamma$.
It also leads to the same form of the BT asymmetry as that for $1/2^+$:
\be
D_{xz}=\sin\theta_{c.m.}
\mbox{Re}\{C_1 C_3^* -C_2 C_4^* +\cos\theta_{c.m.}
(C_1 C_4^*-C_2 C_3^*) \} .
\ee
Quite remarkably, 
the behaviour of $D_{xz}$ due to these two different parities 
now becomes more transparent since the role played by 
the dynamics has been isolated out. Nevertheless, it also 
results in vanishing asymmetries at $\theta_{c.m.}=0^\circ$ and $180^\circ$.


%


The most important difference between the $1/2^-$ and $1/2^+$ cases is that 
the role  of $C_1$ may not be as significant as $f_1$ in the 
production of $1/2^+$ due to the absence of the contact term in
the production of $1/2^-$ state. 
In the Born limit, the main contribution
to  $C_1$, though dominant,
comes from the {\it s}- and {\it u}-channel, which differs from 
the Kroll-Ruderman contribution to $f_1$ in $1/2^+$
production.

In Fig.~\ref{fig:(3)}, the BT asymmetries are calculated 
at $W=2.1$ and 2.5 GeV in the Born limit and including the $K^*$ exchanges. 
Interestingly, a rough $\sin\theta_{c.m.}$ behavior as the $1/2^+$ production
still appears near threshold, but with different sign. 
A detailed analysis shows 
that near threshold, the term 
of $C_1 C_3^*$ is still the dominant one in the BT asymmetry.
Here, the dominant 
contribution to $C_1$ is from the {\it s}-channel, while 
the dominant contribution to $C_3^*$ is from 
the {\it t}-channel kaon exchange
via the decomposition ${\bf q}\cdot\veps_\gamma=
\vsig\cdot{\bf q}\vsig\cdot\veps_\gamma 
+i\vsig\cdot(\veps_\gamma\times\hat{\bf k}){\bf q}\cdot\hat{\bf k}
-i\vsig\cdot\hat{\bf k}{\bf q}\cdot(\veps_\gamma\times\hat{\bf k})$.
Therefore, we have
\be
C_1 C_3^* \simeq
-e_0^2g_{\Theta NK}^2 \frac{\kappa_n}{2M_n}\frac{2|{\bf k}||{\bf q}|}
{(W-M_n)(t-M_K^2)} F_t(k, q) F_s(k, q) ,
\ee
which will be  negative since $\kappa_n=-1.91$. 
In comparison with Eq.~(\ref{approx-1}),
it gives a dynamical reason for the sign difference 
between these two parities.
Meanwhile, the dominance of $C_1 C_3^*$ only holds near threshold. 
With the increasing energy, other terms, e.g. $C_2 C_4^*$, and 
other mechanisms, such as $K^*$ exchange, 
can easily compete against $C_1 C_3^*$
and produce deviations from the $\sin\theta_{c.m.}$ behavior
as shown by the asymmetries at $W=2.5 $ GeV.

\section{Discussions and summaries}

In summary, we have analyzed the double polarization asymmetry,
$D_{xz}$, in $\gamma n\to \Theta^+ K^-$, and showed it
to be a useful filter for determining the parity
of $\Theta^+$, provided its spin-parity is either $1/2^+$ or $1/2^-$.
Due to dynamical reasons, asymmetry $D_{xz}$
near threshold would exhibit a similar behaviour but opposite 
sign. 
The advantage of studying polarization observables is that 
uncertainties arising from the unknown form factors 
can be partially avoided in a phenomenology. 
Therefore, although better knowledge of the form factors
will improve the quantitative predictions, 
it should not change the threshold behaviour of $D_{xz}$
dramatically. 
However, special caution should be given to the roles played 
by a possible spin-$3/2$ partner in the {\it u}-channel, as well as 
{\it s}-channel nucleon resonances. In particular, 
as studied by Dudek and Close~\cite{d-c}, the spin-3/2 partner 
may have a mass close to the $\Theta^+$. Thus, a 
significant contribution from the spin-3/2 pentaquark state
may be possible. Its impact on the BT asymmetry needs to be investigated.
In brief, due to the lack of knowledge in this area, 
any results for the BT asymmetry would be extremely important 
for progress in gaining insights into the nature of pentaquark states
and dynamics for their productions~\cite{zc}. Experimental facilities 
at Spring-8, JLab, ELSA, and ESRF should have access to the BT asymmetry
observable.

\section*{Acknowledgments}

The author thanks Jim Al-Khalili and Frank Close for 
collaborations on the relevant works. 
Useful discussions with Volker Burkert and Takashi Nakano on experimental 
issues, and with Kim Maltman 
and Bingsong Zou on theoretical points are gratefully acknowledged.
Financial supports of the U.K. EPSRC (Grant No. GR/R78633/01 and GR/M82141) 
are acknowledged.

\end{document}